\documentclass[aps,twocolumn,showpacs,prb,floatfix,superscriptaddress]{revtex4-1}

\usepackage[pdftex]{graphicx}
\usepackage{amsmath}
\usepackage{amsfonts}
\usepackage{amssymb}
\usepackage{units}
\usepackage[english]{babel}
\usepackage{acronym}

\renewcommand{\vec}[1]{\boldsymbol{#1}}
\newcommand{\Tr}{\mathrm{Tr}}
\renewcommand{\Im}{\mathrm{Im}}

\begin{document}
\begin{sloppy}

\title[Correlation analysis of spin-dependent transport]{Correlating transmission and local electronic 
structure in planar junctions: A tool for analyzing spin-dependent transport calculations}

\author{P. Bose}
\email[Corresponding author. Electronic address:\ ]{pbose@mpi-halle.de}
\affiliation{Max-Planck-Institut f\"ur Mikrostrukturphysik, Weinberg
  2, D-06120 Halle (Saale), Germany}

\author{P. Zahn}
\affiliation{Institut f\"ur Physik, Martin-Luther-Universit\"at
Halle-Wittenberg, D-06099 Halle (Saale), Germany}

\author{I. Mertig}
\affiliation{Max-Planck-Institut f\"ur Mikrostrukturphysik, Weinberg
  2, D-06120 Halle (Saale), Germany}
\affiliation{Institut f\"ur Physik, Martin-Luther-Universit\"at
Halle-Wittenberg, D-06099 Halle (Saale), Germany}

\author{J. Henk}
\affiliation{Max-Planck-Institut f\"ur Mikrostrukturphysik, Weinberg
  2, D-06120 Halle (Saale), Germany}

\date{\today}

\begin{abstract}
We propose to correlate transmittance maps and spectral-density maps of planar junctions, in order to analyze quantitatively and in detail  spin-dependent transport calculations. Since spectral-density maps can be resolved with respect to atom, angular momentum, and spin, the resulting correlation coefficients reveal unequivocally, e.\,g., which layers or which orbitals determine the tunnel conductances. Our method can be used for transport calculations within the Landauer-B\"uttiker formalism. Its properties and features will be discussed by means of a pure bcc Fe(001) lead as well as an extensively studied Fe(001)/MgO/Fe(001) planar tunnel junction.
\end{abstract}


\newacro{TMR}{tunnel magnetoresistance}
\newacro{TER}{tunnel electroresistance}
\newacro{2BZ}{two-dimensional Brillouin zone}
\newacro{MTJ}{magnetic tunnel junction}
\newacro{KKR}{Korringa-Kohn-Rostoker}
\newacro{PDOS}{partial density of states}
\newacro{TDOS}{total density of states}


\maketitle

\section{Motivation}
Spin electronics---or spintronics for short---is one of the major topics in contemporary physics (see for example Refs.~\onlinecite{Grundler02} and \onlinecite{Wolf06}). 
With respect to both device applications and fundamental physics, planar  junctions have been and are being investigated experimentally and theoretically with great effort. The spin-dependent transport properties of magnetic tunnel junctions show up as  \ac{TMR}, that is the change of the conductance upon reversal of the magnetization direction in one of the two electrodes\cite{Maekawa02,Zutic04}. Replacing the insulating barrier by a ferroelectric, a \ac{TER} effect can be observed in addition. In this case, the conductance depends as well on the orientation of the electrical polarization in the ferroelectric\cite{Tsymbal06}.

In \ac{TMR} experiments, the current that is flowing through a tunnel device is detected in dependence on
 external fields and device parameters (e.g.\ bias voltage and individual layer thicknesses). The current-voltage characteristics of tunnel devices are often interpreted within the Julli\`{e}re model\cite{Julliere75}, 
perhaps due to its simplicity. The validity of this model has been severely questioned\cite{MacLaren97}
because it relates the \ac{TMR} ratio exclusively with the spin polarization of the electrodes and, thus, neglects the interface and barrier regions completely. If the Julli\`{e}re model would be valid, the tunnel conductance would not depend on the interface material at all, in contrast to observations. In other words, the interface region is essential and has to be described in theory as good as possible.

From the preceding it is evident that a reliable description of transport in planar junctions must capture the essential properties on an atomistic level. Hence, present theoretical investigations of transport rely on sophisticated first-principles approaches to the electronic and magnetic structures.

Advanced first-principles approaches to transport allow a very detailed analysis of the electronic structure and the conductance. In particular, the probably most important question---which orbitals in which layer determine the transport properties---can be answered. However, the amount of output data that is produced by modern computer
codes becomes often unmanageable for human beings. Therefore, one restricts oneself to representative subsets or comprises the numerical data into manageable representations. For example, one puts the focus of the analysis on a single wavevector in the \ac{2BZ}, typically to the \ac{2BZ} center\cite{Butler01}.
By doing so, one should be aware that such a restriction could fail because considerable parts of the \ac{2BZ} may contribute to the conductance. Examples for data representations are transmittance and spectral-density maps in the \ac{2BZ} (introduced in Section~\ref{sec:theoretical_analysis}) which have become established analysis tools. They allow in principle to answer the above question. Because they are compared visually, they leave space for speculation and interpretation; or phrased differently, they introduce ambiguity. Apparently, there is need for an improved analysis tool which allows to analyze transport properties unequivocally and quantitatively, rather than ambiguously and qualitatively.

In this paper, we propose a quantitative analysis of transport properties which goes beyond the approaches sketched above. We propose to correlate transmittance and spectral-density maps. The resulting correlation coefficients reveal unequivocally, for example, which layers or which orbitals determine the tunnel conductances. Our method can be used for any computer code which relies on a Landauer-B\"uttiker-type approach, thus being applicable in most of the present-day transport calculations. Its properties will be discussed for two junctions exhibiting a planar geometry.

The paper is organized as follows. In Section~\ref{sec:theoretical_analysis} we introduce our approach for analyzing transport calculations. Section~\ref{sec:conductance_calculations} gives a brief overview of our numerical approach. The correlation analysis is applied in Section~\ref{sec:applications} to a pure Fe(001) lead (\ref{sec:Fe001}) and to an Fe(001)/MgO/Fe(001) \ac{MTJ} (\ref{sec:FeMgOFe}). Concluding remarks are given in Section~\ref{sec:conclusions}.

\section{Theoretical analysis of the tunnel conductance}
\label{sec:theoretical_analysis}
We consider a planar junction which consists of a left electrode $\mathcal{L}$, an interface region $\mathcal{I}$, and a right electrode $\mathcal{R}$. Due to the translational invariance parallel to the interface region, the Bloch states in the electrodes are indexed by the (in-plane) wavevector $\vec{k}_{\parallel}$ in the \ac{2BZ}; $\vec{k}_{\parallel}$ is conserved in the scattering process.

The bias voltage $V$ opens an `energy window of transport'; the chemical potentials $\mu_{\mathcal{L}}$ and $\mu_{\mathcal{R}}$ of $\mathcal{L}$ and $\mathcal{R}$,  respectively, differ by $e V = \mu_{\mathcal{L}} - \mu_{\mathcal{R}}$. Without loss of generality we consider the case $V > 0$, for which incoming occupied Bloch states in $\mathcal{L}$ can be transmitted into outgoing unoccupied Bloch states in $\mathcal{R}$.

According to Landauer and B\"uttiker\cite{Buettiker85,Imry99}, the conductance $C$ is given by
\begin{align}
 C(V) & = \frac{e^{2}}{h}
\int_{\mu_{\mathcal{R}}}^{\mu_{\mathcal{L}}}
\int_{\mathrm{2BZ}}
T^{\mathcal{L} \to \mathcal{R}}(V; E, \vec{k}_{\parallel})\,\mathrm{d}^{2}\vec{k}\,\mathrm{d}E.
\end{align}
The wavevector integral is over the \acl{2BZ}. The transmittance $T^{\mathcal{L} \to \mathcal{R}}(V; E, \vec{k}_{\parallel})$ is the sum over the transmission probabilities of all incoming occupied states $\lambda$ in $\mathcal{L}$ and outgoing unoccupied states $\rho$ in $\mathcal{R}$. It is related to the scattering matrix $S$ of the interface region by
\begin{align}
 T^{\mathcal{L} \to \mathcal{R}}(V; E, \vec{k}_{\parallel})
 & = \sum_{\lambda \rho} \left| S_{\lambda \rho}^{\mathcal{L} \to \mathcal{R}}(V; E, \vec{k}_{\parallel}) \right|^{2}.
 \label{eq:transmission}
\end{align}

The above transmittance is a key quantity in the proposed analysis; it is conveniently displayed versus $\vec{k}_{\parallel}$ at fixed $E$ and $V$, in so-called transmittance maps ($T$ maps). Note that $T$ can be regarded as a global quantity since it depends on the electronic structure of the entire junction, that is both electrodes (incoming and outgoing Bloch states) and the interface region (scattering matrix).

The local electronic structure of the device is described in terms of the spectral density
\begin{align}
 N_{\alpha}(V; E, \vec{k}_{\parallel})
 & = -\frac{1}{\pi} \Im\, \Tr_{\alpha} G^{+}(V; E, \vec{k}_{\parallel}).
 \label{eq:SD}
\end{align}
$\alpha$ is a compound index which can comprise for example layer, atom, orbital, angular momentum, spin indices or point-group representation. 
For a given division of the complete index space, the 
subsets of $\alpha$ are disjoint. The trace $\Tr_{\alpha}$ of the Green function $G^{+}$ is restricted to the given $\alpha$ and, thus, allows a detailed analysis of the electronic structure. As for the transmittance, the spectral density is displayed in maps versus $\vec{k}_{\parallel}$ ($N$ maps).

In previous publications we have analyzed the conductance by relating pronounced features in a transmittance map with features in the associated spectral-density maps. If these features show up in both the $T$ map and in an $N_{\alpha}$ map we concluded that the chosen set $\alpha$ determines the transmittance in this $(V,E)$ region. This way we could show that for Fe/Mn/vacuum/Fe junctions the topmost Mn layer governs their \ac{TMR} ratio\cite{Bose07}.
 However, for more complicated systems it turned out that the visual inspection of the maps becomes ambiguous, tedious, and not very reliable. As a consequence, we propose to correlate properly normalized $T$ and  $N$ maps by means of a projection (inner product). This procedure results in a set of a few unambiguous numbers which allow to determine rapidly the set of significant $\alpha$ indices.

For given energy $E$ and bias voltage $V$ we define the average value of a function $X(V; E, \vec{k}_{\parallel})$ of the transmittance $T(V; E, \vec{k}_{\parallel})$ and the spectral density $N_{\alpha}(V; E, \vec{k}_{\parallel})$ as the average over the \ac{2BZ},
\begin{align}
 A_{X}
 & \equiv
 \frac{1}{\Omega_{\mathrm{2BZ}}}
\int_{\mathrm{2BZ}}
  X(V; E, \vec{k}_{\parallel})\
\,\mathrm{d}^{2}\vec{k},
\end{align}
where $\Omega_{\mathrm{2BZ}}$ is the area of the \ac{2BZ}\footnote{The considered area $\Omega$ has not necessarily to be extended over the whole 2BZ. Principally, $\Omega$ can be restricted to arbitrary parts of it.}. The correlation coefficient $c_{\alpha}(V; E)$ is then defined by
\begin{align}
 c_{\alpha}(V; E)
 & \equiv
 \frac{A_{T N_{\alpha}}}{\sqrt{A_{T^{2}} A_{N_{\alpha}^{2}}}}.
\end{align}
This quantity can be interpreted as an inner product of the normalized $T(V; E, \vec{k}_{\parallel})$ and $N_{\alpha}(V; E, \vec{k}_{\parallel})$. Since the latter are semi-positive for all $\vec{k}_{\parallel}$, $0 \leq c_{\alpha}(V; E) \leq 1$.

Note that $c_{\alpha}(V; E)$ is invariant with respect to scaling $T(V; E, \vec{k}_{\parallel})$ and $N_{\alpha}(V; E, \vec{k}_{\parallel})$ (explicitly $T(V; E, \vec{k}_{\parallel}) \to \tau T(V; E, \vec{k}_{\parallel})$ and $N_{\alpha}(V; E, \vec{k}_{\parallel}) \to \nu N_{\alpha}(V; E, \vec{k}_{\parallel})$).
Further, for a visual inspection and comparison of the $X(V; E, \vec{k}_{\parallel})$ on the same scale it is convenient to normalize
them by
\begin{align}
  \widetilde{X}(V; E, \vec{k}_{\parallel})
  & \equiv \frac{X(V; E, \vec{k}_{\parallel})}{\sqrt{A_{X^2}}}.
  \label{eqn:normX}
\end{align} 

The correlation coefficient, alternatively expressable as $c_{\alpha}(V; E) = A_{\widetilde{T} \widetilde{N}_{\alpha}}$, is a measure for the `overlap' of $T$ and $N_{\alpha}$ or of $\widetilde{T}$ and $\widetilde{N}_{\alpha}$. (i) Consider a constant transmittance and a constant spectral density, $T(V; E, \vec{k}_{\parallel}) = t$ and $N_{\alpha}(V; E, \vec{k}_{\parallel}) = n_{\alpha}$, which can be viewed as `completely overlapping'. Then $A_{T^{2}} = t^{2}$,  $A_{N_{\alpha}^{2}} = n_{\alpha}^{2}$, and $A_{T N_{\alpha}} = t n_{\alpha}$. Consequently, $c_{\alpha} = 1$ which we will consider as perfect correlation. (ii) Consider a $T(V; E, \vec{k}_{\parallel})$ which is nonzero only in a region $\Omega_{T}$ of the \ac{2BZ}:
\begin{align}
T(V; E, \vec{k}_{\parallel})
& = T(V; E, \vec{k}_{\parallel}) 1_{\Omega_{T}}(\vec{k}_{\parallel}),
\end{align}
with the indicator
\begin{align}
  1_{\Omega_{T}}(\vec{k}_{\parallel})
  & = 
  \begin{cases}
  1 & \vec{k}_{\parallel} \in \Omega_{T}
  \\
  0 & \vec{k}_{\parallel} \not \in \Omega_{T}
 \end{cases}.
\end{align}
Likewise, $N_{\alpha}(V; E, \vec{k}_{\parallel})$ is assumed nonzero in a region $\Omega_{N_{\alpha}}$ which is disjoint with $\Omega_{T}$ ($\Omega_{T} \cap \Omega_{N_{\alpha}} = \varnothing$; `zero overlap'). Consequently, $A_{T^{2}} \not= 0$, $A_{N_{\alpha}^{2}} \not= 0$, and $A_{T N_{\alpha}} = 0$, giving $c_{\alpha} = 0$. We consider this case as perfectly uncorrelated. We note in passing that there may be other definitions of correlation coefficients, as it is the case for the correlation of random variables\cite{Bickel77}.

Finally, the \acl{PDOS} \acs{PDOS}$_{\alpha} = A_{N_{\alpha}}$ and the 
\acl{TDOS} \acs{TDOS} $= \sum_{\alpha}A_{N_{\alpha}}$ ($\equiv~A_{N_{\Sigma}}$) are computed
based on $N_{\alpha}$, respectively.

\section{Conductance calculations}
\label{sec:conductance_calculations}
Because our theoretical approach to spin-dependent tunneling has been described 
in detail elsewhere\cite{Tusche05,Bose07,Bose08,Khan08,Bose2010}
we restrict ourselves to a brief survey here. The electronic structure of a tunnel junction
is computed within the local spin-density approximation to density-functional theory, 
as is formulated in multiple-scattering theory\cite{Zabloudil05}. 
Our spin-polarized relativistic layer \ac{KKR} method\cite{omni, Henk02} 
provides on the one hand the Green function of the entire system, from which 
the spectral densities $N_{\alpha}(V; E, \vec{k}_{\parallel})$ can be computed, 
eq.~(\ref{eq:SD}). On the other hand, the Green function can be used to calculate 
the transmission $T(V; E, \vec{k}_{\parallel})$, eq.~(\ref{eq:transmission}), within 
the framework of the Landauer-B\"uttiker theory\cite{Henk06, Saha08}. We can
also follow the approach introduced by MacLaren and Butler\cite{MacLaren99}
in which the Bloch states in the electrodes and the scattering matrix of the interface 
region are computed by means of layer-\ac{KKR} algorithms\cite{MacLaren89a}. 
Here, the Green function is not computed explicitly.

The $T$ and $N$ maps have been computed on identical $\vec{k}_{\parallel}$
meshes in the entire \ac{2BZ}, with at least $40\,000$ points.
In the following applications we restrict ourselves to the case of zero bias ($V = 0$
and $\mu_{\mathcal{L}} = \mu_{\mathcal{R}}$).

\section{Applications}
\label{sec:applications}

\subsection{Fe(001)}
\label{sec:Fe001}

The properties of the proposed correlation analysis are best introduced by an---admittedly---trivial case: the spin-resolved Sharvin 
conductance\cite{Sharvin65} 
of Fe(001). Because the three regions of the junction---$\mathcal{L}$, $\mathcal{I}$, and $\mathcal{R}$---are identical, the scattering matrix in eq.~(\ref{eq:transmission}) is $S_{\lambda\rho} = \delta_{\lambda\rho}$ and $T(\vec{k}_{\parallel})$ is an integer. Accordingly, the $N$ map is a projection of the Fermi surface onto the (001) plane. Without spin-orbit coupling, spin is a good quantum number; hence, we treat the majority and minority channels separately.

\begin{figure*}
 \centering
 \includegraphics[width=0.95\textwidth]{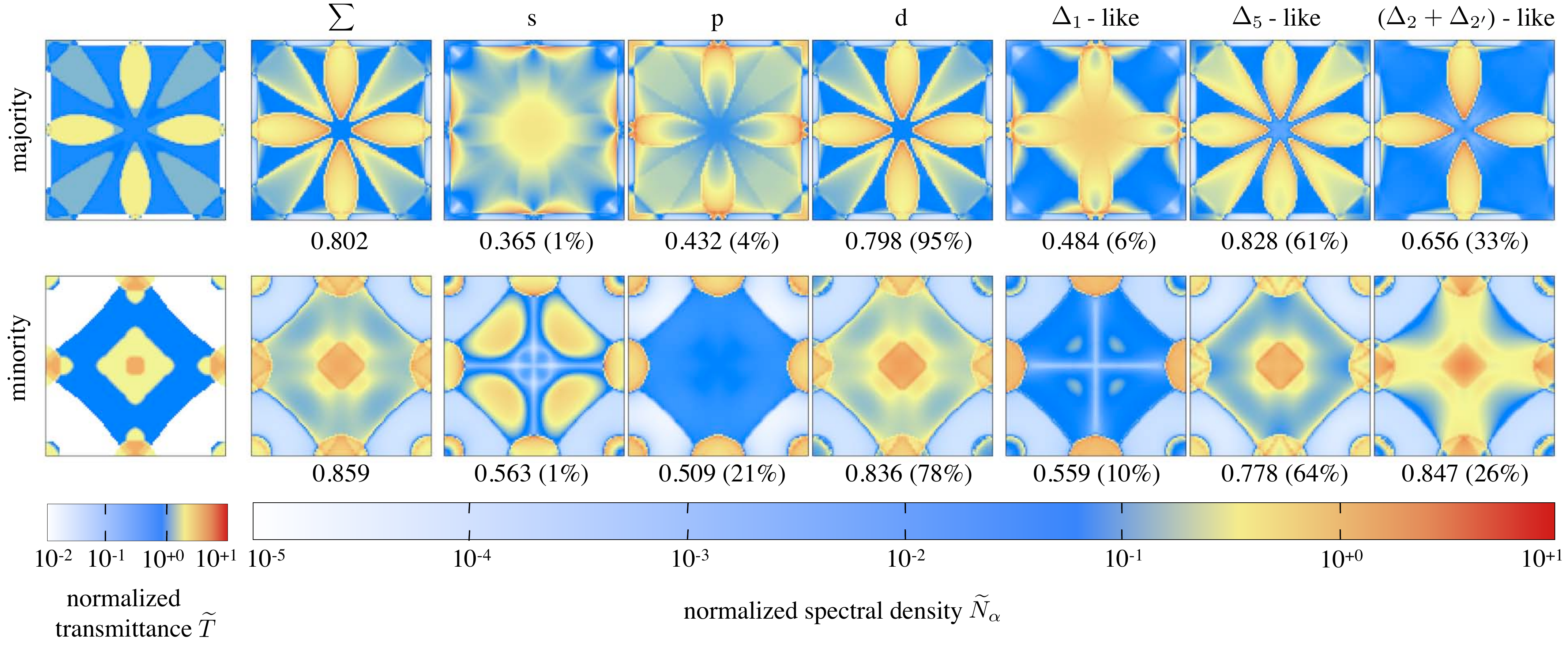}
 \caption{Correlation analysis of the spin-dependent Sharvin conductance of Fe(001). Spin-resolved transmittance and spectral density maps are shown for the entire \ac{2BZ} (top row: majority channel; bottom row: minority channel).  Each map is normalized according to Eq.~\eqref{eqn:normX}. The $\widetilde{N}_{\Sigma}$ maps are decomposed with respect to the representations shown in Table~\ref{tab_decomposition}. The correlations $c_{\alpha}$ are given beneath each $\widetilde{N}_{\alpha}$ map. Numbers in brackets represent each map's partial contribution to the \acl{TDOS}  $\nicefrac{A_{\widetilde{N}_{\alpha}}}{A_{\widetilde{N}_{\Sigma}}}$.  Logarithmic color scales for $\widetilde{T}$ and $\widetilde{N}_{\alpha}$ are given at the bottom.}
 \label{fig:Fe001}
\end{figure*}

Figure~\ref{fig:Fe001} displays the normalized $\vec{k}_{\parallel}$-resolved transmittance and spectral density maps. 
According to Table~\ref{tab_decomposition}, the
later are ordered based on their \{s, p, d\} or $\{\Delta_{1}, \Delta_{5},  \Delta_{2}+\Delta_{2'}\}$-like orbital contributions. 
\begin{table}
  \caption{Decomposition $\alpha$ of the spectral density with respect to angular momentum $(\ell)$ and
                 orbital $(m)$ quantum numbers according to Eq. \eqref{eq:SD}. Incrementing $\ell$ provides
                 a classification by means of s, p, and d orbitals.  An alternative decomposition can be obtained 
                 with respect to the irreducible representations of the point group $C_{4\mathrm{v}}$ 
                 [\onlinecite{Inui90}]. 
                 Because such 
                 a decomposition holds strictly speaking only for the \ac{2BZ} center 
                 ($\overline{\Gamma}$, $\vec{k}_{\parallel} = 0$), we refer to `$\Delta_{1}$-like' maps etc.} 
  \begin{tabular}{c | c | c | c}
    \hline\hline
          $\alpha$      &    $\ell$ &  $m$ & orbitals \\ \hline
         $\Delta_1$-like    & 0, 1, 2   &     0    & s, p$_z$, d$_{3z^2-r^2}$\\
         $\Delta_5$-like    & 1, 2       & -1, 1  & p$_x$, p$_y$, d$_{3xz}$, d$_{3yz}$\\
         $\Delta_2$-like    & 2           & -2      & d$_{x^2-y^2}$\\
         $\Delta_{2'}$-like & 2           & 2       & d$_{3xy}$\\
      \hline\hline
  \end{tabular} 
  \label{tab_decomposition} 
\end{table}
The assignment of the angular momentum orbitals to the different subsets $\alpha$
is motivated by the irreducible representations of the $C_{4\mathrm{v}}$ symmetry group.
This decomposition is given in Table~\ref{tab_decomposition}.

(i) The map of the total spectral density agrees nicely with the associated transmittance
map in the majority channel (top row). In particular, all features are present and the
correlation is consequently sizable ($c_{\Sigma} = 0.802$).

(ii) Other maps which essentially capture  all features within the majority channel are the
$\widetilde{N}_{\mathrm{d}}$ and $\widetilde{N}_{\mathrm{\Delta_{5}\mathrm{-like}}}$ maps.
A visual comparison of both with the transmittance could lead to the conclusion that the 
former matches slightly better than the latter. The correlation analysis, however, shows that
this may be a misinterpretation; both coefficients indicate sizeable correlations ($c_{\mathrm{d}} = 0.798$ and 
$c_{\Delta_5\mathrm{-like}} = 0.828$) but that of the $\Delta_{5}\mathrm{-like}$ subset is slightly larger.

(iii) As a result of the relatively small correlation coefficients of s ($c_{\mathrm{s}} = 0.365$) 
and p ($c_{\mathrm{p}} = 0.432$)  we identify the d majority states as the dominating conducting
channels. Further, with the help of Table~\ref{tab_decomposition} and the $c_{\alpha}$ of the $\Delta$-like
maps, a hierarchy of conducting d states can be 
specified. Since $c_{\Delta_1\mathrm{-like}} = 0.484$ is quite small, the d$_{3z^2-r^2}$ states
seem to play no significant role. Due to the large $c_{\Delta_5\mathrm{-like}}$ the d$_{3xz}$ and d$_{3yz}$ orbitals appear to 
form the leading transport channels, followed  by the d$_{x^2-y^2}$ and d$_{xy}$ states
($c_{(\Delta_2+\Delta_2')\mathrm{-like}} = 0.656$).

(iv) Similar observations can be made for the minority channel (bottom row). Again, the
d states represent the main conducting channels. But in comparison to the majority channel,
the role of (d$_{3xz}$, d$_{3yz}$) and  (d$_{x^2-y^2}$, d$_{3xy}$) orbitals is interchanged.

This example shows that the correlation analysis provides a powerful analysis tool which on one hand fits nicely to 
the visual interpretation of $T$ and $N$ maps. On the other hand, it clearly reveals possible misinterpretations, as
has become evident in point (ii).

With  94\% (majority) and 
66\% (minority) the d states constitute by far the main contributions to the \ac{TDOS}.
In this example the  hierarchy of correlation coefficients often reflects the hierarchy of
partial contributions to the \ac{TDOS} (see Fig.~\ref{fig:Fe001}). This ordering becomes incorrect for instance
in the case of $\Delta_{5}$-like and $(\Delta_{2}+\Delta_{2'})$-like states in the
minority channel. Here, the $\Delta_{5}$-like states represent with 64\% most of the \ac{TDOS},
but exhibit with a $c_{\Delta_5\mathrm{-like}}$ of 0.788  a smaller correlation
than the $(\Delta_{2}+\Delta_{2'})$-like contributions (26\%, $c_{\Delta_2+\Delta_{2'}} = 0.847$).

This last finding indicates already that a one-to-one mapping of \ac{PDOS} hierarchies
and correlation coefficients is not viable. In general, transmittances depend not only
on the number of available states but rely also on other conditions like e.g. the wavefunction matching
at interfaces. Hence, as a prototype  which exhibits more complicated
correlations between transport and electronic structures properties, Fe/MgO/Fe \acp{MTJ} will be 
discussed now.
 
\subsection{Fe(001)/MgO/Fe(001)}
\label{sec:FeMgOFe}

In the following an Fe(001)/MgO/Fe(001) \ac{MTJ} comprising six monolayers
MgO is analysed. The magnetic directions within both Fe leads are collinearly aligned to
each other and considered for the case of a parallel magnetic configuration. Below we
discuss the correlation analysis of the majority channel in more detail, because the
current is dominated by the spin-up carriers\cite{Butler01}.

 The corresponding transmittance map is displayed in Fig.~\ref{fig:maj-transmittance}. 
\begin{figure}[b]
 \centering
 \includegraphics[width = 0.6\columnwidth]{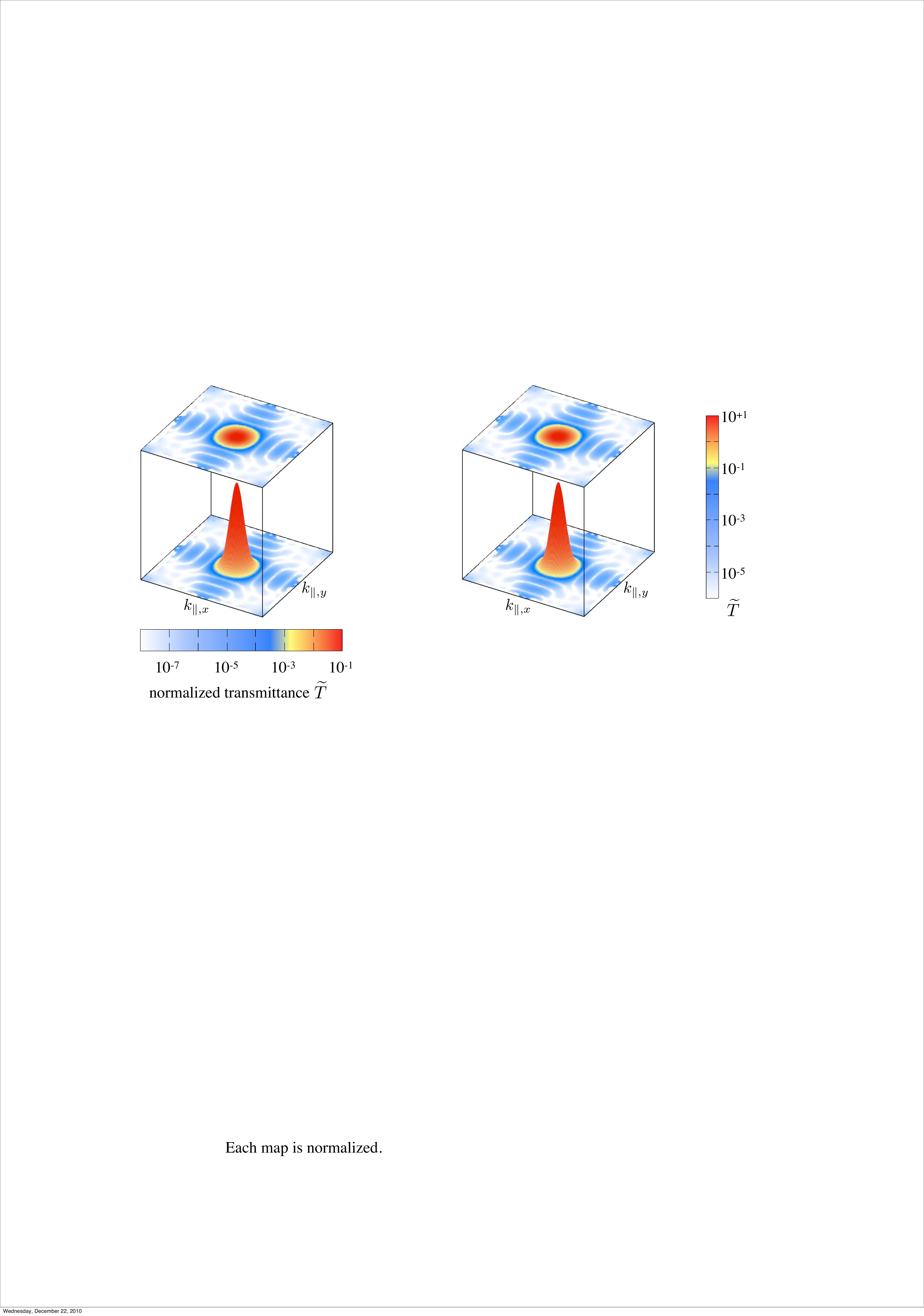}
 \caption{Normalized majority transmittance map $\widetilde{T}(\vec{k}_{\parallel})$ of an Fe(001)/6MgO/Fe(001) \ac{MTJ}.
 The color scale for $\widetilde{T}$ is logarithmic.}
 \label{fig:maj-transmittance}
\end{figure}
As typical  for $\widetilde{T}(\vec{k}_{\parallel})$ in
the majority channel [\onlinecite{Butler01}] a Gaussian-like, 
radially symmetric distribution is found around the \ac{2BZ} center. 

Let us suppose briefly that the 
Julli\`{e}re\cite{Julliere75} 
model is valid and that the transport properties 
can be interpreted exclusively based on the available \acs{PDOS}$_{\alpha}$. Then the
spin-polarized conductances are estimated with the product of the densities of states at the Fermi energy,
strictly speaking with the \acs{PDOS}$_{\alpha}$ inside the left $(\mathcal{L})$ and right lead $(\mathcal{R})$. 
In this picture, the conducting channels would be, according to the previous discussion of iron,
mainly determined by d-like or $\{\Delta_{5}, \Delta_{2}+\Delta_{2'}\}$-like Bloch states 
of the Fe electrodes. But a visual inspection and comparison with the majority 
maps in Fig.~\ref{fig:Fe001} reveals at first glance no similarity in the structures with neither the 
$\widetilde{T}(\vec{k}_{\parallel})$ nor the $\widetilde{N}_{\mathrm{d}}(\vec{k}_{\parallel})$ maps. 

Consequently, one could ask which states are essential for the transport processes if the predominant
Fe bulk states do not play a decisive role. A closer look at the $\widetilde{N}_{\mathrm{s}}$ and 
$\widetilde{N}_{\mathrm{\Delta_{1}\mathrm{-like}}}$ maps in Fig.~\ref{fig:Fe001} leads to the
identification of centrosymmetric blobs like that in Fig.~\ref{fig:maj-transmittance}. 
But the associated Bloch states exhibit small $c_{\alpha}$ and marginal contributions to the \ac{TDOS}
in Fe bulk. Do these states play a decisive role for the transport within the \ac{MTJ}?

\begin{figure}
 \centering
 \includegraphics[width = \columnwidth]{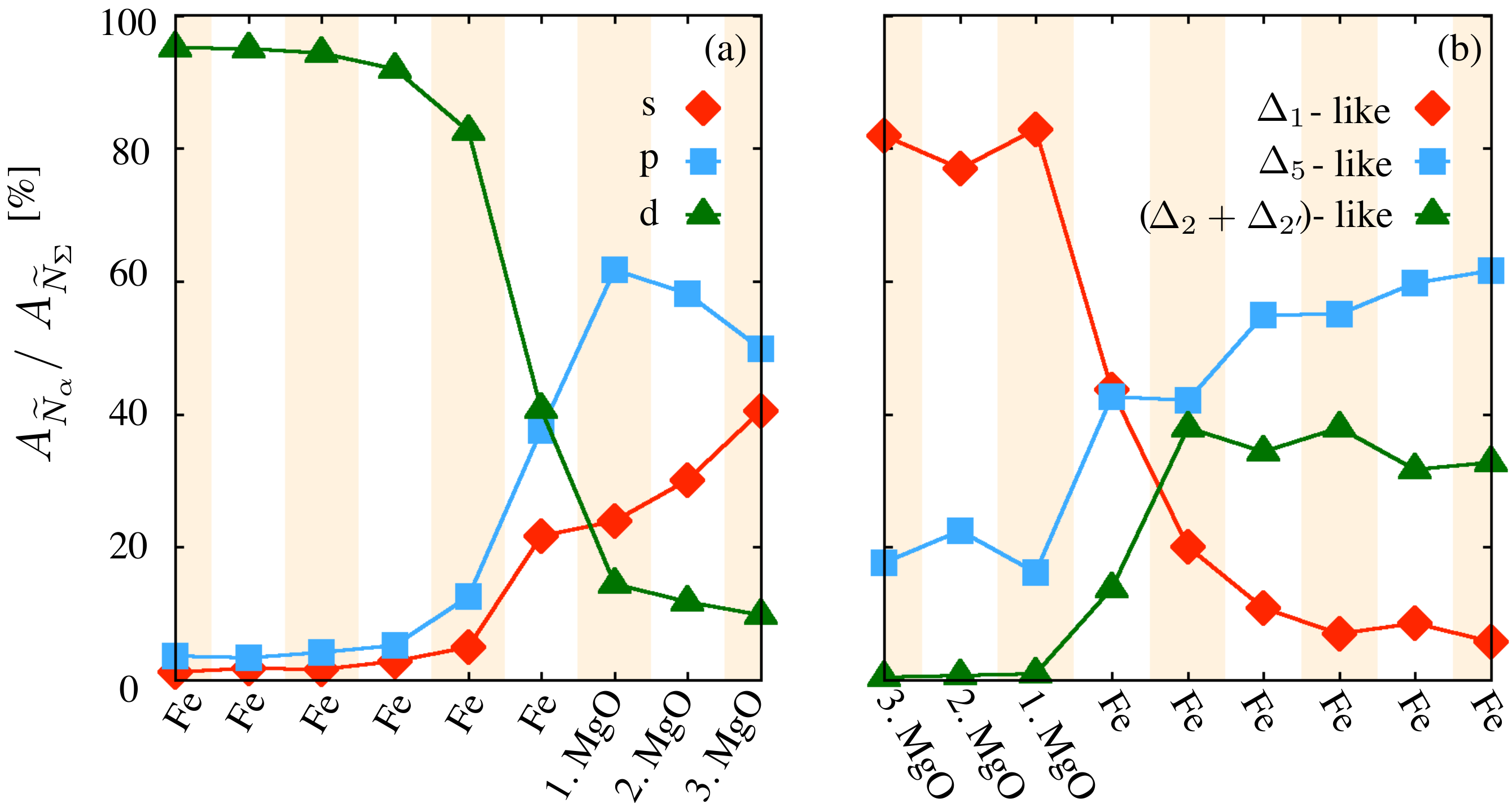}
 \caption{(Color online) Layer- and orbital-resolved contributions of the partial density of states 
                $A_{\widetilde{N}_{\alpha}}$ to the total density of states $A_{\widetilde{N}_{\Sigma}}$ within the
                majority channel of an Fe(001)/6MgO/Fe(001) \ac{MTJ}. The fractions are shown 
                for (a) s, p, d, and  (b) $\{\Delta_{1}, \Delta_{5},  \Delta_{2}+\Delta_{2'}\}$-like orbital decompositions.}
 \label{fig:PDOS-contribution}
\end{figure}
To answer this question we consider in a next step the layer-wise
contributions of the \acs{PDOS}$_{\alpha}$ to the respective \ac{TDOS} at the
Fermi energy (see Fig.~\ref{fig:PDOS-contribution}). In particular, the 
$\nicefrac{A_{\widetilde{N}_{\alpha}}}{A_{\widetilde{N}_{\Sigma}}}$ percentages are shown
for one half of the symmetric \ac{MTJ} and are classified again by means of
\{s, p, d\} or $\{\Delta_{1}, \Delta_{5},  \Delta_{2}+\Delta_{2'}\}$-like orbital decompositions. 

The values for the outermost Fe layers in Fig.~\ref{fig:PDOS-contribution} (a) and (b) are
identical to those for the pure Fe lead in Fig.~\ref{fig:Fe001}, indicating
the bulk-like character of those layers far from the interfaces. The corresponding d states represent
with 80\%-95\% the most numerous parts of the \ac{TDOS} up to the second Fe layers adjacent to the
MgO interfaces. At these interfaces the number of d states is drastically reduced. Within the
MgO film the decrease is continued down to about 10\% inside the middle region of the
tunnel barrier. 

On the other hand, the s and p contributions which are apparently vanishingly small
within the Fe electrodes, obtain substantial weight inside the MgO spacer.
In particular, the p states exhibit a maximal percentage of 60\% within the 1. MgO
monolayer. Deeper inside the MgO, this fraction reduces to about 40\%. In contrast,
the s fractions reach a level of 20\% at the Fe/MgO interfaces and increase monotonously
to roughly about 40\%.  Thus, in the middle of the MgO the number of p and s states are comparably large,
with a slight advantage of the former. 

Further, a decomposition in $\{\Delta_{1}, \Delta_{5},  \Delta_{2}+\Delta_{2'}\}$-like contributions
reveals completely different characteristics in Fig.~\ref{fig:PDOS-contribution}b. Here, the $\Delta_{1}$-like
contributions, which are of minor importance within the Fe leads, experience a massive increase at
the Fe/MgO interfaces and reach levels of about 80\% within the tunnel barrier. The remaining 
20\% are predominantly occupied with $\Delta_{5}$-like and few $(\Delta_{2}+\Delta_{2'})$-like
states. This hierarchical order of \acs{PDOS}$_\alpha$ within the MgO reflects the well known fact of
symmetry-selective decay lengths of the evanescent Bloch states within the tunnel 
barrier\cite{Butler01, Heiliger06b}.

Due to their sizable presence within the bottle neck of the junction, i.~e. the MgO barrier,
s and $\Delta_{1}$-like states might indeed characterize the transport.
But whether this predominance also results in a dominance of the conducting
channels can only be answered by an analysis of the transmittance map in 
Fig.~\ref{fig:maj-transmittance} with the respective $\widetilde{N}_{\alpha}$ 
maps. In order to specify exactly which layer-resolved spectral densities fits best the structures 
of the transmittance map, the $\widetilde{N}_{\alpha}$ maps
of the whole Fe/MgO/Fe tunnel junction have to be inspected.
These maps are shown in Fig.~\ref{fig:FEMGOcorrelation-maps}a.

\begin{figure*}
\centering
 \includegraphics[width = 0.8\textwidth]{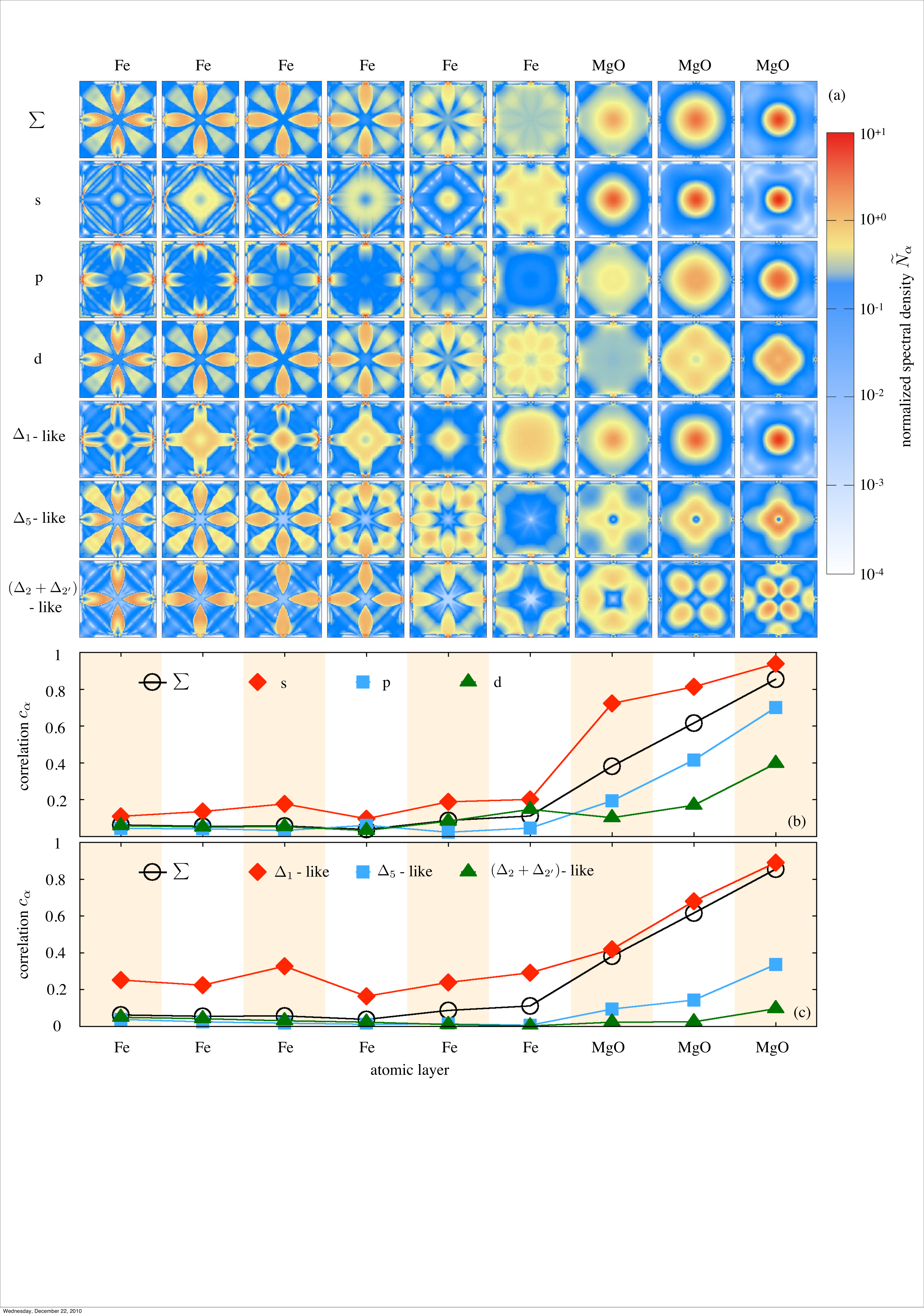}
 \caption{Correlation analysis for the majority channel of an Fe(001)/6MgO/Fe(001) 
  \ac{MTJ} in parallel magnetic configuration.  (a) Layer- and orbital-wise ordered spectral density maps 
  $\widetilde{N}_{\alpha}$ (left $\rightarrow$ right) are shown for the entire \ac{2BZ} with decompositions $\alpha$
   according to Table~\ref{tab_decomposition} (top $\rightarrow$ bottom). Correlation coefficients 
  $c_\alpha$ for  contributions (a) with $\alpha =$ s, p, d   and  (b) $\alpha = \Delta_{1}$-like, 
  $\Delta_{5}$-like, $(\Delta_{2}+\Delta_{2'})$-like.  Color scales for $\widetilde{N}_{\alpha}$ are logarithmic.}
 \label{fig:FEMGOcorrelation-maps}
\end{figure*}

One can see immediately that $\widetilde{N}_{\Sigma}$, $\widetilde{N}_{\mathrm{s}}$,
$\widetilde{N}_{\mathrm{p}}$, and $\widetilde{N}_{\Delta_1\mathrm{-like}}$ within
the MgO layers exhibit a great similarity with the structure of $\widetilde{T}(\vec{k}_{\parallel})$
in  Fig.~\ref{fig:maj-transmittance}. On the other hand, the similarities of $\widetilde{N}_{\mathrm{d}}$,
$\widetilde{N}_{\Delta_5\mathrm{-like}}$ and $\widetilde{N}_{(\Delta_2+\Delta_{2'})\mathrm{-like}}$
within the same layers appear rather small. Due to the great dissimilarities of the spectral density maps within the Fe layers it is 
reasonable to expect low correlations there. But, based on visual comparisons it
is hard to make definitive statements about similarities of the structures
and hence which states provide the largest contributions to the electronic
transport of the whole MTJ.

At this point, the tool of the correlation analysis provides the potential to gain
clearer statements. The computed correlation coefficients which
represent a measure for the similarity of two maps are summarized
in Fig.~\ref{fig:FEMGOcorrelation-maps}b and \ref{fig:FEMGOcorrelation-maps}c.

(i) Considering Figure~\ref{fig:FEMGOcorrelation-maps}b, it turns out that s 
states exhibit the most prominent correlations in all layers.
For each Fe layer the $c_{\mathrm{s}}$ can be quantified
with about 0.1-0.2 as nearly double as large as those of p and d states. Together
with the significantly high correlations of 0.7 (1.~MgO layer) up to 0.95 (3.~MgO layer)
inside the tunnel barrier, the previously assumed dominant role of the s states
can be regarded as proven for the entire MTJ.

(ii) However, inside the MgO layers the p and d states exhibit sizeable
increases of their correlation coefficients, too. In particular, the  
$c_{\mathrm{p}}$ are with 0.2 (1.~MgO layer), 0.4 (2.~MgO layer) and 0.7 (3.~MgO layer)
approximately twice as high as the $c_{\mathrm{d}}$. In the
discussion of the $c_{\Delta_{1}\mathrm{-like}}$ coefficients below it
will become evident that the increasing $c_{\mathrm{p}}$ and 
$c_{\mathrm{d}}$ are mainly related to states exhibiting p$_{\mathrm{z}}$ and
d$_{3z^2-r^2}$ orbital character.

(iii) Since the \ac{TDOS} maps comprise 
structures of s states closley correlated to the transmission map  and less correlated p and d states,
the correlation values of $c_{\Sigma}$ are always lower than those for the s states.

(iv) Along the atomic layers of the MTJ, the characteristics of the $c_{\Delta_{1}\mathrm{-like}}$ coefficients in
Fig.~\ref{fig:FEMGOcorrelation-maps}c show 
a qualitatively similar dominance as it was found for the $c_{\mathrm{s}}$ coefficients in Fig.~\ref{fig:FEMGOcorrelation-maps}b. Within
the Fe electrodes the correlations of the $\Delta_{1}$-like states are just slightly but noticeable
larger than those of the s states  in these layers, indicating an additional relevance of 
p$_{\mathrm{z}}$ and d$_{3z^2-r^2}$ orbitals in the electronic transport. This
finding is substantiated by relatively small correlations of both other $\Delta$-like representations 
inside the tunnel barrier. Since the latter exhibit only \{p$_x$, p$_y$, d$_{3xz}$, d$_{3yz}$\}
and \{d$_{x^2-y^2}$, d$_{xy}$\} orbital character (see Table~\ref{tab_decomposition}),
it is reasonable to assume that the rising importance of p and d states inside the MgO is identical 
to an increasing significance of p$_{\mathrm{z}}$ and d$_{3z^2-r^2}$ orbitals.

The comparison of the results of the correlation analysis with the discussion of 
available PDOS$_{\alpha}$ in Fig.~\ref{fig:PDOS-contribution} shows common
features but reveals also significant differences. A common outcome of
both approaches is the principal importance of the tunnel barrier for the
electronic transport of the MTJ. In particular, both discussions end up
with a conclusion that $\Delta_{1}$-like states--- i.~e. s, p$_z$ and d$_{3x^2-r^2}$
orbitals which are preferentially aligned along the transport direction--- tunnel
most effectively and consequently carry the dominant part of the tunnel current.

The fact, that $\Delta_{1}$-like states, especially those with s orbital character, show their
decisive role also within the Fe leads, represents a qualitative different outcome of the 
correlation analysis. In contrast, these states exhibit the lowest PDOS$_{\alpha}$
contributions within the Fe layers in Fig.~\ref{fig:PDOS-contribution}. 

In principle, the symmetry-selective filtering of the MgO tunnel barrier
shows up  by means of the hierarchical order of the \{$\Delta_{1}, \Delta_{5},
\Delta_{2}+\Delta_{2'}$\}-like PDOS$_{\alpha}$ fractions in Fig.~\ref{fig:PDOS-contribution}.
But deeper within the MgO these percentages stay rather constant. The layer-wise increase
of the effective tunneling processes of $\Delta_{1}$-like states inside
the MgO can only be seen by the increasing characteristics of the $c_{\Delta_{1}\mathrm{-like}}$ 
coefficients in Fig.~\ref{fig:FEMGOcorrelation-maps}c.

\section{Conclusions and remarks}
\label{sec:conclusions}

The analysis of transport properties within contacts that exhibit planar
geometries is often accompanied with a visual comparison of large ensembles of 
$\vec{k}_{\parallel}$-resolved local spectral density and transmission maps. In this article we presented an
analysis tool which helps to avoid laborious inspections and potential
misinterpretations by providing exact measures. Applying the proposed 
correlation analysis the contributions of atoms or orbitals can be quantified
unambigously. The dominance of the s states in the tunnel current
of Fe/MgO/Fe \acp{MTJ} could be proven even for the states in the Fe leads,
where the contribution of states to the \ac{TDOS} is only marginal. 

Further, the method extends the popular discussion of transport proporties at the $\bar{\Gamma}$
point in Fe/MgO/Fe \acp{MTJ} to a more comprehensive anlysis which comprises the entire \acl{2BZ}. 
In principle, the technique is not restricted to planar junctions. Beside more complex 
layered structures it can be applied to other nano structures like point contacts or nano wires.
For these atomic-sized contacts, we expect that the corresponding correlation coefficients provide
the same physical insights as one would obtain from an analysis of the conduction eigenchannels [\onlinecite{Bagrets07}].

\begin{acknowledgments}
This work is supported by the DFG's \textit{Sonderforschungsbereich}
762 `Functionality of Oxide Interfaces'.
\end{acknowledgments}

\end{sloppy}

\bibliographystyle{apsrev4-1}
\bibliography{papers}

\end{document}